\documentclass[twocolumn,twoside,slac_two]{revtex4}
\usepackage{graphicx}
\usepackage{fancyhdr}
\newcommand{\arcsec}{$^{\prime\prime}$\/}

\begin{document}

\title{Coordinated Fermi/Optical Monitoring of Blazars and the Great 2009 September Gamma-ray Flare of 3C 454.3}


\author{P. S. Smith, E. Montiel, S. Rightley, J. Turner, G. D. Schmidt}
\affiliation{Steward Observatory, University of Arizona, Tucson, AZ 85721, USA}
\author{B. T. Jannuzi}
\affiliation{NOAO, Kitt Peak National Observatory, Tucson, AZ 85726, USA}

\begin{abstract}
We describe the optical spectropolarimetric monitoring program at Steward
Observatory centered around gamma-ray-bright blazars and the LAT Monitored
Source List planned for Fermi Cycles 2--4.  The large number of measurements
made during Cycle 1 of the Fermi mission are available to the research 
community and the data products
are summarized (see http://james.as.arizona.edu/$\sim$psmith/Fermi).  The optical
data include spectropolarimetry at a resolution of $\sim$20~\AA,
broad-band polarization and flux measurements, and flux-calibrated spectra 
spanning 4000-7600~\AA. These data provide a comprehensive view of the optical
variability of an important sample of objects during the Fermi Era.
In addition to broad-band flux and linear polarization monitoring, the
spectra allow for the tracking of changes to the spectral index of the
synchrotron continuum,
importance of non-synchrotron emission features, and how and when the
polarization varies with wavelength, an important clue as to the structure
of the emission region or the identification of multiple nonthermal components. 
As an illustration, we present observations of 3C 454.3 obtained
in 2009 September during an exceptionally bright gamma-ray flare.
The blazar was optically bright during the flare, but except for a few
short periods, it showed surprisingly low polarization ($P < 5$\%).
Opportunities exist within the Fermi research community to coordinate
with our long-term optical monitoring program toward the goal of maximum 
scientific value to both the Fermi and associated radio VLBI monitoring
of blazars. 

\end{abstract}

\maketitle

\thispagestyle{fancy}


\section{INTRODUCTION}

The high apparent luminosity and rapid variability of $\gamma\/$-ray
emission observed in blazars~\cite{kniffen}.
lead to straightforward
physical arguments implying that, regardless of the mechanism producing
the $>$MeV photons, the gamma radiation is relativistically
beamed as is the bulk of the lower-energy emission~\cite{urry}.
This realization
presents the important opportunity during the era of the Fermi
Gamma-ray Space Telescope ({\it Fermi\/})
to examine for the first time if a direct physical
connection can be made
between the sites of $\gamma$-ray production and those
generating the beamed, polarized optical synchrotron
continuum in these active galactic nuclei (AGNs).
The program at Steward Observatory (SO), described here, is designed
to test for this physical connection by providing a unique
set of optical linear polarization and flux spectra of blazars.

A fundamental connection between the source(s) of $\gamma$-rays
and polarization in AGNs is already inferred
given that all blazars detected at high energy also show
large and similarly variable optical continuum polarization.
In addition, polarization measurements provide the only direct means of
gaining information on
the magnetic field within the synchrotron emission region.
The level of linear polarization of a blazar gives a
measure of the degree of ordering
of the magnetic field within the emission region, and this can be very
high (fractional polarization $P \gtrsim 40$\%).
The polarization position angle yields the
orientation of the field responsible for the
polarized flux projected on the sky.

Despite the evidence amassed suggesting
that the bulk of the observed emission comes from a
single source - the relativistic jet - it has been generally difficult
to find systematic correlations between the variations seen in flux and
polarization of different spectral regions
(or even within a single spectral region),
making it hard to use variability
for constraining models of the continuum emission regions.
However, optical polarimetry combined with VLBI observations
have provided insights about the physical structure and processes involved
in generating the powerful and rapidly variable emission in 
blazars~\cite{gabuzda}, \cite{darcangelo}, \cite{marscher1}.

In the past,
it has been nearly
impossible to study possible relationships between
$\gamma\/$-ray and lower-energy emission in AGNs because of the
difficulty in obtaining systematic, high signal-to-noise
ratio ($S/N\/$) monitoring of sources
at high energies.
The first year of the {\it Fermi\/} mission has changed this situation with the
Large Area Telescope (LAT)
demonstrating the capability to
monitor the 0.1--300~GeV emission for a large sample
of AGNs on time scales as short as a day~\cite{abdo1}, \cite{abdo2}.

\section{THE OBSERVATIONAL PROGRAM}

\subsection{Scope of the Monitoring Campaigns}

The optical program is designed to nightly
monitor $\gamma\/$-ray-bright blazars,
both in linear polarization and flux, for about one week every month
and to release the data obtained to the research community
as quickly as possible.
This density of observation provides a good match to the $\gamma\/$-ray
variability data that {\it Fermi\/} can provide for the brightest
blazars in the sky, allowing for the often rapid $\gamma\/$-ray variations to
be directly compared to both the polarization and flux behavior
observed for the optical synchrotron continuum.
Ideally, optical polarization and flux monitoring should be a continuous
as possible, but
the number of nights per year devoted to this intensive monitoring
at Steward Observatory is constrained by lunar phase, competition
from other programs for telescope time, and the increases in manpower 
that would be required to significantly expand the duration and/or 
frequency of the
observing campaigns.
An additional consideration is the substantial effort
needed to fully process the 
polarization, flux, and 
spectral data described in \S3 so that the results publicly
presented are definitive.
Indeed, an observing campaign uninterrupted by bad weather presents a 
formidable challenge to fully reduce before the start of the next campaign.
The optical program started in 2008 October and is planned to continue until
at least the end of Cycle~4 of the {\it Fermi\/} mission (2012 August).

During a typical night $\sim$20 objects can be observed spectropolarimetrically
with the 2.3--1.5~m telescopes available for use.  
The core of the object sample is the LAT-monitored list of blazars chosen for
Cycle~1 of the {\it Fermi\/} mission
(http://fermi.gsfc.nasa.gov/ssc/data/policy/LAT\_ Monitored\_Sources.html).
This sample has been augmented by blazars that have flared in $\gamma\/$-rays
during the first year and a half of the mission and announced in Astronomer's
Telegrams (ATELs; http://www-glast.stanford.edu/cgi-bin/pub\_rapid)
as interesting targets of opportunity.
Table~1 listed the blazars that have been observed during the first 11
observing campaigns of the SO program.

\begin{table}[t]
\begin{center}
\caption{Blazars Observed during the First 11 Observing Campaigns}
\begin{tabular}{|l|c|r|c|}
\hline \textbf{Object} & \textbf{\# Obs.} & \textbf{$\Delta P\/$ (\%)} &
\textbf{$\Delta V\/$}
\\
\hline 3C 66A & 21 & 7.5--19.7 & 14.05--14.60 \\
\hline AO 0235+164 & 75 & 0.6--23.5 & 15.67--18.79 \\
\hline PKS 0420$-$014 & 12 & 5.8--11.9 & 16.58--17.21 \\
\hline PKS 0454$-$123 & 17 & 0.2--7.3 & $\cdots$ \\
\hline PKS 0528+134 & 16 & 0.8--12.9 & 19.60--20.87 \\
\hline S5 0716+714 & 36 & 2.4--15.5 & 13.13--14.00 \\
\hline 4C 14.23 & 10 & 4.4--20.3 & $\cdots$ \\
\hline PKS 0735+178 & 10 & 1.6--6.4 & 16.21--16.45 \\
\hline OJ 248 & 56 & 0.1--1.6 & 17.31--17.54 \\
\hline OJ 287 & 82 & 11.6--35.5 & 14.01--15.63 \\
\hline Mrk 421 & 72 & 0.6--6.1 & 13.05--13.62 \\
\hline PKS 1118$-$056 & 15 & 1.1--22.5 & $\cdots$ \\
\hline H 1219+305 & 31 & 1.8--7.5 & 16.31--16.60 \\
\hline W Com & 63 & 3.3--17.4 & 14.28--15.31 \\
\hline PKS 1222+216 & 8 & 0.8--4.4 & 15.60--15.90 \\
\hline 3C 273 & 43 & 0.1--1.5 & 12.66--12.84 \\
\hline PKS 1244$-$255 & 19 & 0.6--6.9 & $\cdots$ \\
\hline 3C 279 & 49 & 1.7--34.5 & 15.31--16.97 \\
\hline PKS 1406$-$076 & 6 & 8.0--17.4 & $\cdots$ \\
\hline H 1426+428 & 31 & 0.2--1.8 & 16.50--16.82 \\
\hline PKS 1502+106 & 29 & 5.2--45.2 & $\cdots$ \\
\hline PKS 1510$-$08 & 42 & 1.4--22.6 & 15.13--17.04 \\
\hline B2 1633+382 & 45 & 0.6--7.8 & $\cdots$ \\
\hline 3C 345 & 26 & 0.8--9.5 & 16.23--17.39 \\
\hline Mrk 501 & 67 & 0.1--5.6 & 13.84--14.03 \\
\hline GB6 B1700+6834 & 9 & 1.9--11.2 & $\cdots$ \\
\hline NRAO 530 & 6 & 2.8--6.3 & $\cdots$ \\
\hline 1ES 1959+650 & 40 & 3.2--8.5 & 14.61--15.02 \\
\hline PKS 2155$-$304 & 45 & 0.9--10.1 & 12.87--13.91 \\
\hline BL Lac & 86 & 4.2--25.6 & 14.67--15.39 \\
\hline CTA 102 & 11 & 0.3--1.7 & 17.08--17.21 \\
\hline 3C 454.3 & 140 & 0.2--13.8 & 14.54--16.52 \\
\hline 1ES 2344+514 & 59 & 0.5--3.4 & $\cdots$ \\
\hline
\end{tabular}
\label{l2ea4-t1}
\end{center}
\end{table}

\subsection{Facilities and Instrumentation}

The SO optical program utilizes either the 2.3~m Bok Telescope on 
Kitt Peak, AZ (http://james.as.arizona.edu/$\sim$psmith/90inch.html),
or the 1.54~m Kuiper Telescope on Mt. Bigelow, AZ
(http://james.as.arizona.edu/$\sim$psmith/61inch).
Given that the Kuiper Telescope was designed foremost for planetary 
and lunar observations, its mount does not allow the telescope
to be operated at Declinations north of $\sim +$61$^{\circ}$, whereas there
is no such restriction with the Bok Telescope. 

All observations are obtained using the SPOL spectropolarimeter
\cite{schmidt}.
SPOL is a versatile, high-throughput ($\sim$30\% instrument+telescope),
moderate resolution ($R \sim 300$--1000),
dual-beam spectropolarimeter that employs a waveplate and Wollaston prism
to modulate and analyze polarized light.  
A $\lambda$/4 waveplate is inserted into the telescope beam just after 
the slit to detect circular polarization,
a $\lambda$/2 waveplate is used to measurement linear polarization.
The instrument is also able to provide imaging polarimetry over a narrow
field of view ($\sim 50 \times 50$~arcsec$^2$ at both 2.3~m and 1.5~m
telescopes) with the substitution of
a diffraction grating by a plane mirror in the light path. 
The detector is a $1200 \times 800$~pixel$^2$, thinned,
antireflection-coated SITe CCD having
$\lesssim$2.5~e$^{-}$ read noise.

The intent of the monitoring program is to provide a comprehensive view
(both the flux and polarization {\it spectra\/})
of the optical emission from $\gamma\/$-ray-bright blazars for comparison
with their high-energy emission as well as with VLBI monitoring at radio
wavelengths.
Therefore, the default configuration of SPOL employs a 600~l~mm$^{-1}$ 
grating, providing spectral coverage of 4000--7550~\AA\ 
in first order, with a resolution
of 15--25~\AA, depending on the slit width chosen for the observation.
Seven slits, each with a length of $\sim$50\arcsec , provide a range
of widths from 1\arcsec -- 13\arcsec .
Typically, slit widths of 2\arcsec -- 3\arcsec 
are used for 
spectropolarimetry depending on the seeing
and the brightness of the target, while the widest slits are
utilized for the differential
spectrophotometry described in \S3.2.

Occasionally, SPOL is configured for imaging polarimetry 
during an observing campaign.
This is usually done on marginal nights when observations need
to be completed more quickly
during short breaks in the cloud cover.
In these cases, the spectral coverage of the measurements are defined
by a standard Johnson $V\/$ filter bandpass.

\subsection{Calibration}

Polarization measured by SPOL is rotated to the conventional astronomical 
reference frame by observing known interstellar polarization standard
stars~\cite{lupie}.
At least two standards are observed per campaign as a consistency check.
The uncertainty in the absolute calibration of the polarization
position angle ($\theta\/$) from the grid of standards used is not folded into 
the estimated uncertainty of a measurement.
The uncertainties reported for $\theta\/$ are statistical as systematic
errors in the absolute calibration would apply equally to all measurements
within an observing campaign.

The efficiency of the $\lambda\/$/2-waveplate is checked every campaign
by inserting a full-polarizing Nicol prism into the beam and measuring
the result.
The efficiency of the semi-achromatic $\lambda\/$/2-waveplate used in SPOL
varies between 0.97 at $\sim$4500~\AA\ to nearly 1.0 at $\sim$6500~\AA, before
falling back off to 0.97 by $\sim$8000~\AA.
This calibration is applied to all polarization results.

As a check for instrumental polarization, at least one unpolarized standard
star~\cite{lupie} is observed per campaign.
No correction for instrumental polarization is made to the data as SPOL
consistently shows that the polarization imparted by the telescope and
instrument is $\lesssim 0.1$\%.

All calibration of the optical flux measurements is from differential
flux measurements made using stars in the fields of the targeted blazars.
The majority of targets observed since Fermi launch have calibrated
field stars (see e.g.,~\cite{raiteri}, \cite{smithb}, \cite{gonzalez}),
though 10 do not,
and as a result, no range in $V\/$ magnitude
is given for these objects in Table~I.
The difference in magnitude between the blazars and the identified comparison
star is given whenever differential spectrophotometry is performed, so
in principle the apparent magnitude can be recovered in the future if 
the comparison stars are calibrated. 

Optical spectra, usually from the spectropolarimetric
measurements, are flux calibrated by observing several spectrophotometric
standard stars throughout an observing campaign.
An average sensitivity function is derived from these observations to
convert the instrumental spectra to $F_\lambda\/$ with units of
erg~cm$^{-2}$~s$^{-1}$~\AA$^{-1}$ over the range $\lambda = 4000$--7550~\AA.
Corrections are applied to the flux density spectra for the
airmass of the target during the observation and the elevation of
observatory using the standard KPNO extinction curve within the IRAF
reduction package (see \cite{stone}).

\section{THE AVAILABLE DATA PRODUCTS}

All of the reduced data obtained during this program are publicly available at 
http://james.as.arizona.edu/$\sim$psmith/Fermi.

\subsection{Polarimetry}

As of the end of campaign \#11 (2009 November), almost 1300 polarization
measurements of 33 blazars have been made (see Table~I).
The primary data products of the spectropolarimetry are spectra of the
normalized linear polarization Stokes parameters $q\/$ and $u\/$ spanning
4000--7550~\AA.
The median $q\/$ and $u\/$ for $\lambda = 5000$--7000~\AA\ is determined from
the spectra and used to calculate $P\/$ and $\theta\/$.
All four quantities are summarized in tables both sorted by time and by
object, with the 
measurement uncertainties determined from the RMS measured in the 
Stokes spectra.

The full-resolution Stokes spectra are also made available
in addition to the summary tables so that
investigators may bin the spectropolarimetry as desired.
Providing the $q\/$ and $u\/$ spectra is more straightforward than 
producing equivalent spectra of $P\/$ and $\theta\/$ because
of the normal error distribution of the Stokes parameters.
All reported values of $P\/$ have been corrected for statistical bias
inherent in these measurements~\cite{wardle}.
The correction used is 
\begin{equation}\label{eq:bias}
P=\sqrt{(q^2+u^2)-\sigma_p^2}.
\end{equation}
This correction is insignificant given the very high
$S/N\/$ measurements resulting from binning the spectra over
2000~\AA\ (500 pixels), when typically, $\sigma_p < 0.1$\% based on
photon statistics.
However, the measurement bias becomes significant if the spectra are
much more finely binned, or for objects that are faint and/or nearly 
unpolarized.

On occasion, $V\/$-band imaging polarimetry is obtained
instead of spectropolarimetry.
As there are no Stokes spectra produced from these observations,
the polarimetry is only summarized in the data tables.

\subsection{Photometry and Flux Spectra}

For clear nights (or clear portions of nights), the results from differential
spectrophotometry are summarized in nightly photometry tables.
Tables that summarize entire observing campaigns
are also available that sort the photometry by object.
These measurements are independent of the spectropolarimetry as they utilize
the widest slits available in the instrument to avoid variations in seeing
and wavelength-dependent slit losses, especially
when a target is observed at a low elevation in the sky. 
These large-aperture observations yield a flux spectrum of the blazar 
and one or more comparison stars in the field.
After convolving the the flux spectra with a synthetic $V\/$ filter
bandpass, the flux is summed and the magnitude differences 
computed. 

For those fields that have calibrated magnitudes for the 
comparison stars, the apparent $V\/$ magnitude of the blazar
is reported in the photometry summary tables.
For all measurements, the differential magnitude and
the identification of the comparison star(s)
used are reported so that investigators can revise the reported
apparent magnitudes if there are future refinements to the 
calibration of the field stars, or if a currently uncalibrated field
is calibrated in the future.

Since the flux spectra span 4000--7550~\AA, both the standard $V\/$ and
$R\/$ photometric bandpasses are encompassed.
By default, the results at $V\/$ are reported, but the data exist to
construct blazar light curves at $R\/$ as well as any other filter
bandpass covered by the spectra.
Investigators interested in using the differential spectrophotometry
to combine with photometry obtained in filters other than Johnson
$V\/$ should contact the authors to have the appropriate data made
available for use.

Flux spectra resulting from the spectropolarimetry are of much higher
$S/N\/$ than the spectra from the large-aperture spectrophotometry because 
they are the co-addition of 16 exposures taken at various waveplate
positions.
In contrast, the spectrophotometric spectra are from a single exposure with a
much higher sky background because of the wider slit employed.
The spectra from the spectropolarimetry are calibrated using the 
averaged sensitivity function for the observing campaign (see \S2.3)
and made available so that the optical continuum and any spectral
features can be followed and compared 
with the $\gamma\/$-ray flux variations detected by {\it Fermi\/}.

Occasionally, spectra suffer from slit-loss problems when objects
are observed at airmasses of 2 or greater since the slit is not
oriented at the parallactic angle (the slit is left aligned east-west
for all observations so as not to disrupt the calibration of $\theta\/$).
This problem is corrected by closely comparing the continuum shape
of the blazar from the narrow-slit observation with that from the
wide-slit spectrophotometry, which is insensitive to the effects of
differential atmospheric refraction.
The final, reduced spectrum is corrected to match the lower-$S/N\/$, wide-slit
observation.
In addition, the flux spectrum is scaled to the results of the 
synthetic $V\/$-band photometry for that night, although this is
not done on non-photometric nights or for objects with no calibrated
comparison stars.

For each monitored object, the reduced full-resolution total
flux and Stokes ($q\/$ and $u\/$) spectra are available for 
download in a single
gzipped tar file.
As of 2009 August, the spectra are complete for all of the optical
observations obtained during Cycle~1 of the {\it Fermi\/} mission.

\section{3C 454.3 IN 2009 SEPTEMBER}

An example of the results from the SO optical monitoring program
are given by the observations of 3C~454.3 obtained during a 
very active period of $\gamma\/$-ray emission in 2009 September
(optical campaign \#9).
Figure~1 shows the optical measurements over a two-week period,
along with the 0.1--300~GeV flux as measured by {\it Fermi\/}.
Error bars are plotted for all of the data, but these are generally smaller
than the points denoting the optical photometry and polarimetry.
The polarization measurements are determined from the
$\lambda = 5000$--7000~\AA\ median values of $q\/$ and $u\/$.

\begin{figure}[t]
\centering
\includegraphics[width=90mm]{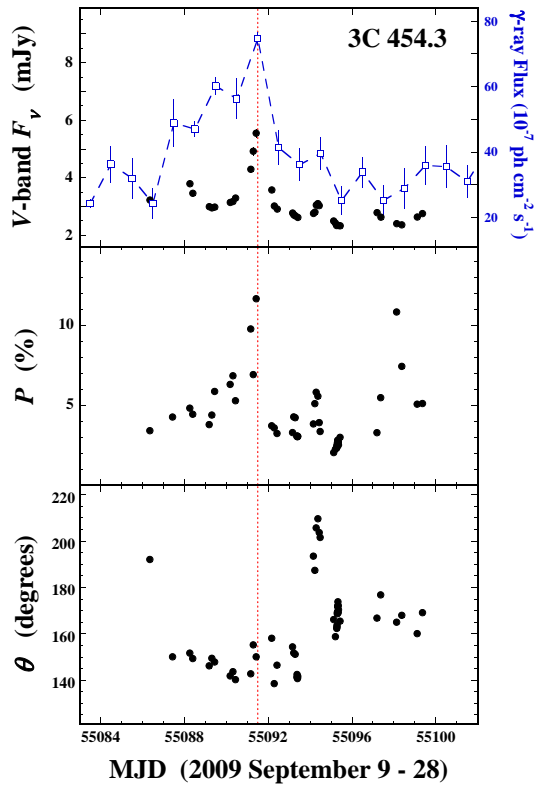}
\caption{Optical flux and linear polarization measurements
of 3C~454.3 obtained during the first major emission peak in the
$\gamma\/$-ray light curve at the start of flaring period for the
blazar in 2009 September.
Gamma-ray fluxes at 0.1--300~GeV are shown in the top panel as
open squares connected by {\it dashed\/} line segments.
The vertical line through all panels denotes the peak $\gamma\/$-ray
flux from this period.
There is a high degree of correlation between the high-energy flux
and both the total and polarized optical fluxes.} \label{JACpic2-f1}
\end{figure}

In the top panel of Figure~1, a strong correlation can be seen between
the $\gamma\/$-ray and optical fluxes, with both reaching a sharp peak
on 2009 Sep~17 (UT).
The correlation with the $\gamma\/$-ray flux is even more pronounced
in polarized flux ($P \times F_\nu\/$).
The polarization generally increases steadily until Sep~17, when 
$P \sim 12$\%, and then drops to $\sim$3\% within a day.
In contrast, the polarization position angle only shows large
differences with its value on Sep~17 ($\theta \sim 150^{\circ}\/$) 
several days prior to and after the peak of the $\gamma\/$-ray flare.

Although the systematic behavior seen in Figure~1 is highly suggestive
of the common region producing the optical and $\gamma\/$-ray photons,
even this relatively short time series of
observations reveals complications.
Note that the high level of polarization observed on MJD~= 55098 is
not accompanied by an increase in $\gamma\/$-ray flux.
Also, large rotations in $\theta\/$ are seen during periods of
relatively  stable optical and $\gamma\/$-ray emission.

Figure~2 shows the full-resolution optical flux and linear polarization
spectra from 2009 Sep~17.
In the top panel, the total flux spectrum of Sep~17 is compared to
the spectrum of 3C~454.3 obtained on 2009 Jan~30, when the blazar was optically
two magnitudes fainter ($V \sim 16.5$)
and was not detected by {\it Fermi\/} on daily
time scales. 
The polarization of the blazar is not compared for these two epochs since 
it was essentially unpolarized ($P < 1$\%) on Jan~30.

\begin{figure}[t]
\centering
\includegraphics[width=90mm]{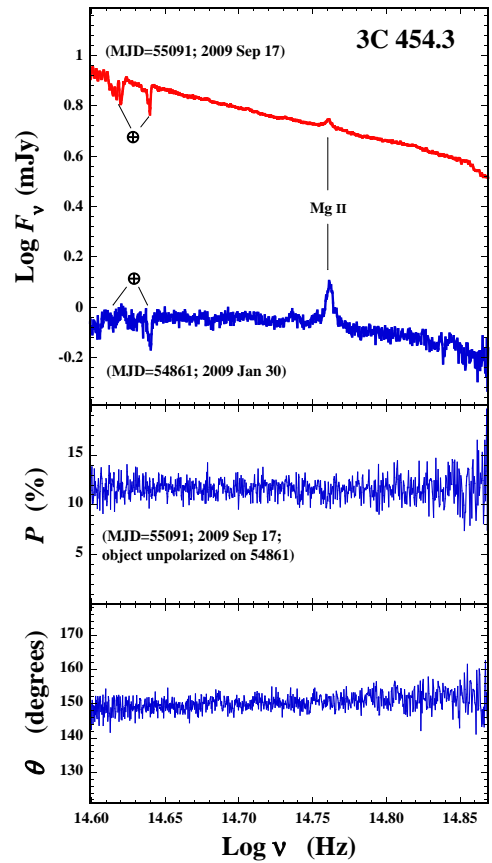}
\caption{Full-resolution optical flux and linear polarization data
obtained on 2009 Sep~17 near the peak of the 0.1--300~GeV flux.
In the top panel, the optically bright flux spectrum is compared
to that from 2009 Jan~30, when 3C~454.3 was much fainter and 
essentially unpolarized.  ({\it Fermi\/} was unable to detect the
blazar daily during much of 2009 Jan/Feb.)
Also notice that
the optical spectrum is much redder when the object is bright,
as the steep synchrotron continuum dominates and nearly washes
out Mg~II.  Spectra are shown in the observed frame.}
\label{JACpic2-f2}
\end{figure}

From Figure~2, it can be seen that the spectroscopic data delivered
by the SO monitoring program adds several pieces of important information
that are not available from the broad-band measurements.
First, the slope of the optical continuum can be measured, which is an
important diagnostic for the optically thin synchrotron source producing
the polarized flux.
Indeed, the spectral index of the polarized flux can be determined 
{\it independently\/} of any contribution to the optical spectrum from
sources such as the Big Blue Bump (BBB), the host galaxy, and emission lines.
Second, the variation with wavelength of the polarization can be measured.
Wavelength-dependent $P\/$ and/or $\theta\/$ is an important diagnostic
for the structure of the emission region or for the identification 
of multiple polarized non-thermal emission regions.
Unpolarized flux components, such as starlight from the blazar host
galaxy and the BBB, often difficult to detect by analysis of the flux 
spectrum alone can be uncovered in the spectrum of $P\/$~\cite{smith1}.
Finally, the strength of emission and absorption features can
be monitored over the duration of the SO program in a systematic fashion.

For the case of 3C~454.3 shown in Figure~2, the differences in the 
flux spectra between bright and faint optical and $\gamma\/$-ray periods
are readily apparent.
The bright optical spectrum is much redder than that of 2009 Jan~30, as
the steep power-law synchrotron spectrum dominates the continuum and
makes the Mg~II emission line much less prominent.
The faint optical spectrum is much bluer, presumably because of the dominance
of the unpolarized BBB found from observations of
the wavelength dependence of $P\/$~\cite{smith2}.
For this particular example, 3C~454.3 was nearly unpolarized on
Jan~30, consistent with the BBB dominating the continuum.
The telltale signature for the BBB ($P\/$ declining in the blue, with
$\theta\/$ constant with $\lambda\/$) is, however, seen throughout the
SO monitoring program in the polarized
spectra of the object during epochs of intermediate brightness. 

\section{UTILIZING DATA FROM THE SO MONITORING PROGRAM}

Researchers are encouraged to contact the Principal Investigator
of the SO monitoring program (P. S. Smith; psmith@as.arizona.edu) if
they have questions or problems using the data in the public
archive for their investigations.
Although collaboration is not a condition for use of these data,
we request that the source of the observations be cited
(and the PI informed by email) in any
publication that utilizes
results from the Steward Observatory
monitoring program. 
The program is reviewed by the Steward Observatory Telescope Allocation
Committee every four months and knowledge of the utility of these
observations in the study of blazars helps our ability
to garner telescope time for the continuation of the project.
Fully collaborative efforts are also encouraged.

\bigskip % extra skip inserted
\begin{acknowledgments}
This work has been made possible by {\it Fermi\/} Guest Investigator
grant NNX08AV65G.
E. Montiel thanks the NASA Space Grant program for an Internship
during 2008--09.
\end{acknowledgments}

\bigskip % extra skip inserted

\end{document}